\documentclass{eas}
\usepackage{graphicx}

\begin{document}

\title
{Magnetic Field Structure from Synchrotron Polarization}

\author{Rainer Beck}

\address{MPI f\"ur Radioastronomie, Auf dem H\"ugel 69,
53121 Bonn, Germany}

\begin{abstract}

Total magnetic fields in spiral galaxies, as observed through their total synchrotron
emission, are strongest (up to $\simeq30~\mu$G) in the spiral arms.
The degree of radio polarization is low; the
field in the arms must be mostly turbulent or tangled. Polarized synchrotron
emission shows that the resolved \emph{regular} fields are generally strongest in
the interarm regions (up to $\simeq15~\mu$G), sometimes forming
``magnetic arms'' parallel to the optical arms. The field structure is spiral
in almost every galaxy, even in flocculent and bright irregular types which lack spiral
arms. The observed large-scale patterns of Faraday rotation in several massive
spiral galaxies reveal coherent regular fields, as predicted by
dynamo models. However, in most galaxies observed so far no simple patterns of
Faraday rotation could be found. Either many dynamo modes are superimposed and
cannot be resolved by present-day telescopes, or most of the apparently regular field
is in fact \emph{anisotropic random\/}, with frequent reversals, due to shearing and
compressing gas flows. In galaxies with massive bars, the polarization pattern
follows the gas flow. However, around strong shocks in bars, the compression of the
regular field is much lower than that of the gas; the regular field decouples from
the cold gas and is strong enough to affect the flow of the diffuse warm gas.  --
The average strength of the total magnetic field in the \emph{Milky Way\/} is
$6~\mu$G near the sun and increases to 20--40~$\mu$G in the Galactic center region.
The Galactic field is mostly parallel to the plane, except in the center region.
Rotation measure data from pulsars indicate several field reversals,
unlike external galaxies, but some reversals could be due to distortions of the
nearby field.

\end{abstract}

\maketitle

\section{The polarization foreground from the Milky Way}
\label{sect1}

CMB experiments critically depend on the correction for emission from the
Galactic foreground. Several surveys of the total Galactic radio emission and
maps of many selected fields exist (Reich et al.\ \cite{R04b}). The first
absolutely calibrated all-sky polarization survey has just been completed at
$\lambda21$~cm (1.4~GHz) (see Wolleben et al.\ \cite{W06} for the northern part)
which shows that polarized intensity has a much more complex structure than total
intensity. It is not known yet which of the polarization structures, invisible
in total emission, are real and which are due to wavelength-- and latitude--dependent
Faraday effects. The existing polarization data are insufficient to construct a realistic
model. One of the ``toy'' models for CMB correction (Giardino et al. \cite{G02})
was based on an analysis of the angular power spectra of $\lambda13$~cm (2.4~GHz)
polarization data in a region near the Galactic plane, with the (unrealistic)
assumption that  polarization angles have a random distribution. Furthermore,
at $\lambda13$~cm Faraday effects are still strong in the Galactic plane.
All-sky polarization surveys at higher frequencies are urgently needed,
as well as better models of the magneto-ionic interstellar medium.

Another approach is to model the Galactic radio emission at high frequencies with
help of models of the distribution of cosmic ray and magnetic fields in the
Milky Way. However, testing the models is difficult. The available Galactic
data from the radio and $\gamma$ ranges and direct cosmic-ray detections are
affected by the Local Bubble and other local structures. Radio observations
of external spiral galaxies help to model the large-scale properties of
regular and turbulent magnetic fields and cosmic rays.

\begin{figure}[htb]
\center
\centerline{\includegraphics[bb = 49 221 545 612,width=9cm,clip=]{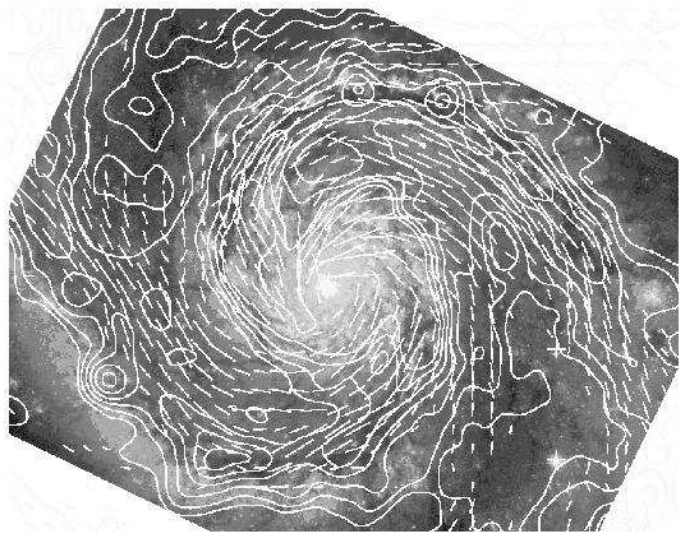}}
\vspace{0.2cm}
\caption{Total radio intensity (contours) and $\mathbf{B}$--vectors of
polarized intensity in the inner disk of the grand design spiral galaxy
M~51 at $6^{\prime\prime}$ resolution, combined from VLA and Effelsberg
observations at $\lambda$6~cm. The field size is $4^{\prime}\times3^{\prime}$.
The length of the vectors is proportional to the polarized intensity.
The background image shows the optical emission observed with the
Hubble Space Telescope (Hubble Heritage Project)
(from \protect{Fletcher et al.\ \cite{F06}}).
}
\label{m51}
\end{figure}

\section{Magnetic fields and synchrotron polarization observations}
\label{sect2}

Learning about magnetic fields is not only important for Galactic models.
Magnetic fields are a major agent in the interstellar medium (ISM). They
contribute significantly to the total pressure which balances the ISM
against gravitation. They may affect the gas flows in spiral arms and
around bars. Magnetic fields are essential for the onset of star
formation as they enable the removal of angular momentum from the
protostellar cloud during its collapse. MHD turbulence distributes
energy from supernova explosions within the ISM. Magnetic reconnection
is a possible heating source for the ISM and halo gas. Magnetic fields
also control the density and distribution of cosmic rays in the ISM.

\emph{Polarized radio emission}, emitted by
cosmic-ray electrons spiralling around magnetic fields in the interstellar
or intergalactic medium, in pulsars and in background quasars, typically
has a much higher degree of polarization than in the other spectral ranges.
The observed degree of polarization is reduced by unpolarized thermal emission
within the beam, which may dominate in star-forming regions,
by \emph{Faraday depolarization\/} (Sokoloff et al.\ \cite{SB98}), and by
geometrical depolarization due to variations of the field orientation within
the beam.

The orientation of the polarization vectors is changed in a magneto-ionic
medium by \emph{Faraday rotation\/} which increases with $\lambda^2$. At small
$\lambda$ (below about 6~cm for typical ISM magnetic fields or
about 1~cm for strong fields in starburst regions) the $\bf{B}$--vectors
(i.e. the observed $\bf{E}$--vectors rotated by $90^{\circ}$) trace,
within a few degrees,
the \emph{orientation\/} of the field in the sky plane (Sect.~\ref{sect5}).
As $\bf{B}$--vectors are ambiguous by $\pm 180^{\circ}$ and hence insensitive
to field reversals, coherent regular and anisotropic random (incoherent) fields
cannot be distinguished. Coherent regular fields and their \emph{direction\/}
can be derived only from the strength and sign of Faraday rotation
(Sect.~\ref{sect6}).

\begin{figure}[htb]
\center
\centerline{\includegraphics[bb = 57 161 552 642,width=8cm,clip=]{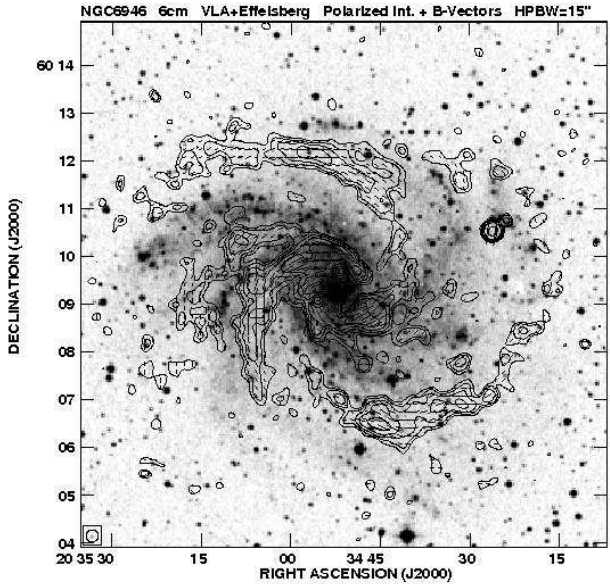}}
\vspace{0.2cm}
\caption{Polarized radio intensity (contours) and $\mathbf{B}$--vectors
of NGC~6946 at $15^{\prime\prime}$ resolution, combined
from VLA and Effelsberg observations at $\lambda$6~cm. The background
grey-scale image shows the optical emission from the DSS
(from \protect{Beck \& Hoernes\ \cite{BH96}})
}
\label{n6946}
\end{figure}

\section{Magnetic field strengths in galaxies}
\label{sect3}

The average strength of the total $\langle B_{\mathrm{t},\perp}\rangle$
and the resolved regular field (or anisotropic random field)
$\langle B_{\mathrm{reg},\perp}\rangle$
in the plane of the sky can be derived from the total and polarized
radio synchrotron intensity, respectively, if energy-density
equipartition between cosmic rays and magnetic fields is assumed
(Beck \& Krause\ \cite{BK05}).

The mean equipartition strength of the total field for a sample of 74
spiral galaxies is $\langle B_\mathrm{t}\rangle\simeq9~\mu$G (Niklas\ \cite{N95}).
Radio-faint galaxies like M\,31 and M\,33 have
$\langle B_\mathrm{t}\rangle\simeq6~\mu$G, while $\simeq15~\mu$G
is typical for grand-design galaxies like M\,51, M\,83 and NGC~6946.
In the prominent spiral arms of M\,51 the total field strength is
$\simeq30~\mu$G (Fletcher et al.\ \cite{F04b}), with peaks along
the prominent dust lane (Fig.~\ref{m51}). The strongest fields in spiral
galaxies were found in starburst galaxies, like M~82 with $\simeq~50~\mu$G
strength (Klein et al.\ \cite{K88}) and the ``Antennae''
(Chy\.zy \& Beck\ \cite{C04}), and in nuclear starburst regions,
like in NGC~1097 (Beck et al.\ \cite{BF05}) and NGC~7552
with $\simeq100~\mu$G strength (Beck et al.\ \cite{BE04}). If energy losses
of electrons are significant, these values are lower limits (Beck \& Krause\
\cite{BK05}).

\begin{figure}[htb]
\center
\centerline{\includegraphics[bb = 57 168 559 649,width=7cm,clip=]{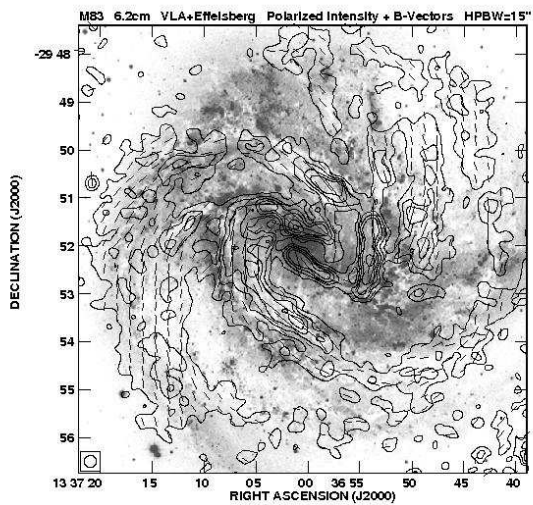}}
\vspace{0.2cm}
\caption{Polarized radio intensity (contours) and $\mathbf{B}$--vectors of
M~83 at $15^{\prime\prime}$ resolution, combined from VLA and
Effelsberg observations at $\lambda$6~cm (Beck, Ehle \& Sukumar, unpublished).
The background optical image is from the Anglo Australian Observatory
(kindly provided by D.~Malin)}
\label{m83}
\end{figure}

The strengths of \emph{regular\/} fields (or anisotropic random fields)
$B_\mathrm{reg}$ in spiral galaxies (observed with a spatial resolution of a few
100~pc) are typically 1--5~$\mu$G.  Exceptionally strong regular fields
are detected in the interarm regions of NGC~6946 of up to $\simeq 13~\mu$G
(Beck \& Hoernes\ \cite{BH96}) and $\simeq15~\mu$G at the inner edge of the
inner spiral arms in M\,51 (Fletcher et al.\ \cite{F04b}). In the spiral arms
of external galaxies the regular field is generally weak (Figs.~\ref{n6946}
and Fig.~\ref{m83}), whereas the turbulent and unresolved tangled fields are
strong due to turbulent gas motions and the expansion of supernova remnants.
Dwarf irregular galaxies with almost chaotic rotation host
turbulent fields with strengths comparable to spiral galaxies, but no
detectable regular fields (Chy\.zy et al.\ \cite{C03}).

\section{Magnetic energy densities}
\label{sect4}

The relative importance of various competing forces in the interstellar
medium can be estimated by comparing the corresponding energy densities.
In the local Milky Way, the energy densities of turbulent gas motions,
cosmic rays, and magnetic fields are similar (Boulares \& Cox\ \cite{BC90}).
Global studies are possible in external galaxies (Beck\ \cite{B04}).
In NGC~6946, the energy density of the magnetic field $E_{magn}$
(assuming energy density equipartition with cosmic rays) is one order of
magnitude larger than that of the ionized gas $E_{th}$
(assuming a volume filling factor of 5\%).
In the inner parts of NGC~6946, the energy densities of the total magnetic
field and of turbulent gas motions are similar, but the field energy
dominates in the outer parts due to the large radial scale length of the
magnetic energy ($\simeq8$~kpc), compared to the scale length of
about 3~kpc for the neutral gas density. If equipartition is no longer
valid in the outer regions due to the lack of cosmic rays, the scale
length of the magnetic field is even larger.

The energy density of global rotation of the neutral gas in the inner part
of NGC~6946 is $\simeq500\times$
larger than that of the turbulent gas motion. As the radial decrease of magnetic
energy density is much slower than that of the rotational energy density,
the magnetic field energy density may reach the level of global rotational
gas motion at some radius and hence affect the rotation curve, as proposed by
Battaner \& Florido (\cite{BF00}). However, this may only happen far out in
a galaxy and is unsufficient explain flat rotation curves.

In the outer parts of galaxies synchrotron emission cannot be detected, but
the strengths of regular fields can still be measured by Faraday rotation of
polarized background sources. Han et al.\ (\cite{H98}) found evidence that the
regular field in M\,31 extends to at least 25~kpc radius without significant
decrease in strength. Hence we may consider that galaxies are surrounded by
huge magnetic halos.

\section{Large-scale magnetic field structures in galaxies}
\label{sect5}

Maps of the total radio emission and maps of the mid-infrared dust
emission of external galaxies reveal a surprisingly close connection
(Frick et al.\ \cite{F01b}, Vogler et al.\ \cite{V05}).
The strongest total fields generally coincide
with highest emission from dust and gas in the spiral arms (Fig.~\ref{m51}).
This suggests a coupling of the tangled magnetic field to the warm dust mixed
with the cool gas (Helou \& Bicay\ \cite{H93}, Niklas \& Beck\ \cite{N97},
Hoernes et al.\ \cite{HB98}).

\begin{figure}[htb]
\center
\centerline{\includegraphics[bb = 64 176 552 634,width=7cm,clip=]{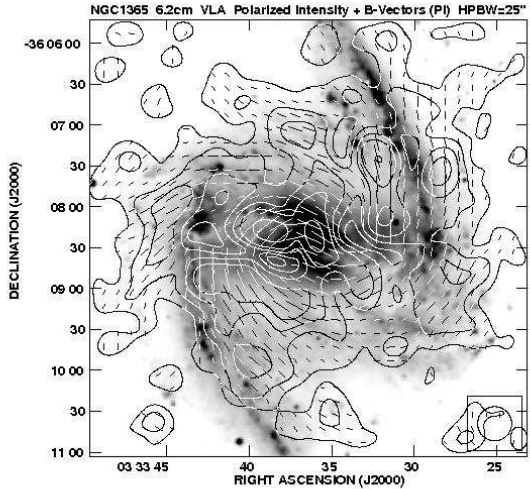}}
\vspace{0.2cm}
\caption{Polarized radio intensity (contours) and $\mathbf{B}$--vectors
of the barred galaxy NGC~1365 at $25^{\prime\prime}$ resolution
from VLA observations at $\lambda$6~cm. The grey-scale image shows an
optical emission from the ESO observatory (kindly provided by P.O.~Lindblad)
(from \protect{Beck et al.\ \cite{BF05}}).
}
\label{n1365}
\end{figure}

The $\mathbf{B}$--vectors of polarized emission form spiral patterns, not only in
spiral galaxies, but also in flocculent and several bright irregular galaxies
(see examples shown in Beck\ \cite{B05}). The magnetic field runs almost parallel
to the spiral arms. In M~51, the field orientation varies by 10$^\circ$--20$^\circ$
around the pitch angles of the spiral arms as traced by CO emission
(Patrikeev et al.\ \cite{P06}). In galaxies with strong density waves,
like M~51, the regular field fills the whole interarm space, but is strongest
at the positions of the dust lanes on the inner edge of the spiral arms
(Fletcher et al.\ \cite{F04b}, Fig.~\ref{m51}). In most other galaxies, the
polarized emission is strongest \emph{between\/} the optical arms, sometimes
forming \emph{magnetic spiral arms\/} with ridge lines between the optical
arms (like in NGC~6946, Fig.~\ref{n6946}, and in M~83, Fig.~\ref{m83}) or
across the arms, like in NGC~3627 (Soida et al.\ \cite{S01}).

In strongly barred galaxies, gas and stars move in highly noncircular orbits.
Numerical models show that gas streamlines are deflected in the bar region along
shock fronts, behind which the gas is compressed in a fast shearing flow.
M83 (Fig.~\ref{m83}) hosts a small bar, NGC~1097 and NGC~1365 (Fig.~\ref{n1365})
host huge bars. The total radio intensity is strongest in the region of the dust lanes,
consistent with compression by the bar's shock. The polarized intensity in NGC~1097
and NGC~1365 (Fig.~\ref{n1365}) has a completely different distribution.
The $\mathbf{B}$--vectors and the gas streamlines around the bar as obtained
in numerical simulations are surprisingly similar, which suggests that the
regular magnetic field is generally aligned with the shearing flow
(Beck et al.\ \cite{B99a}). Remarkably, the optical images of NGC~1097 and
NGC~1365 show dust filaments which are almost perpendicular to the bar and aligned
with the $\mathbf{B}$--vectors. However, the upstream/downstream contrast in polarized
intensity is significantly smaller than that expected from compression and
shearing of the regular magnetic field, which could be the result of decoupling
of the regular field from the dense molecular clouds (Beck et al.\ \cite{BF05}).
The regular field in the bar is probably strong enough to resist shearing
by the rarefied diffuse gas. These observations, for the first time, indicate that
\emph{magnetic forces can control the flow of the diffuse interstellar gas
at kiloparsec scales\/}.

Present-day polarization observations are limited by sensitivity at high
resolution. The best available spatial resolutions are 100--300~pc
in the nearest spiral galaxies and 10~pc in the LMC (Gaensler et al.\
\cite{G05}). Data on beam and Faraday depolarization in NGC~6946 (Beck et al.\
\cite{B99b}), assuming single-sized cells,
indicate that the interstellar field is tangled on scales of
$\simeq20$~pc. The EVLA and the planned Square Kilometre Array (SKA) will
have much improved sensitivities. The SKA will allow to study magnetic
field structures in nearby galaxies at angular resolutions more than 10$\times$
better than today (Beck \& Gaensler\ \cite{BG04}).

\section{Spiral fields: coherent or anisotropic?}
\label{sect6}

Dynamo action generates \emph{coherent\/} spiral magnetic fields, while
compression or shearing of turbulent fields by gas flows generates
\emph{incoherent anisotropic random\/} fields, which reverse direction frequently
within the telescope beam, so that Faraday rotation is small while the degree of
polarization can still be high. (Note that polarization angles are insensitive
to field reversals.) Only Faraday rotation is a signature of coherent regular
fields, and the sense of Faraday rotation reveals the direction of the field.
Large-scale (``mean field'') galactic dynamo models generate spiral fields with
large-scale coherence. Observation of large-scale patterns in Faraday
rotation shows that a significant fraction of the field has a coherent
direction and has been generated by a large-scale dynamo.

The large-scale field structure obtained in $\alpha$--$\Omega$ dynamo models is
described by modes of different azimuthal, radial, and vertical symmetries
(Beck et al.\ \cite{B96}). Several modes can be excited. In flat disks with
axisymmetric gas distribution and smooth rotation, the strongest mode
is the axisymmetric spiral one ($m=0$) with even vertical symmetry (quadrupole),
followed by the weaker $m=1$ (a bisymmetric spiral field), etc.
These modes can be identified by Fourier analysis of their specific
azimuthal RM patterns (Krause\ \cite{K90}).

\begin{figure}[htb]
\center
\centerline{\includegraphics[bb = 57 161 552 656,width=7cm,clip=]{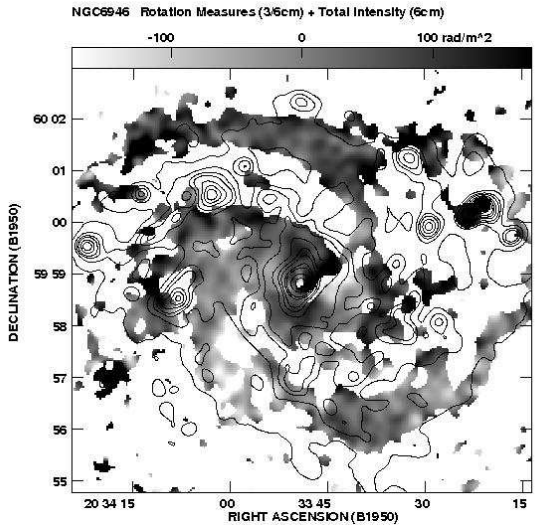}}
\vspace{0.2cm}
\caption{Grey-scale map of the Faraday rotation measures of NGC~6946 between
$\lambda$3~cm and $\lambda$6~cm at $15^{\prime\prime}$ resolution, overlayed
onto contours of total intensity  at $\lambda$6~cm. The data were obtained from
combined polarization observations with the VLA and Effelsberg telescopes
(Beck, unpublished).
}
\label{n6946rm}
\end{figure}

Real galaxies are not uniform. Consequently, recent dynamo models include
the non-axisymmetric gas distribution in spiral arms (Moss\ \cite{M98})
or the gas flow in a bar potential (Moss et al.\ \cite{M01}) where
higher modes may be amplified faster than in the standard model.
Gravitational interaction with another galaxy may also modify the
mode spectrum and enhance the bisymmetric mode (Moss\ \cite{M95}).

M~31 hosts a strongly dominating axisymmetric field (Berkhuijsen et al.\
\cite{BB03}), extending to at least 25~kpc radius (Han et al.\ \cite{H98})
and at least 1~kpc height above the galaxy's plane (Fletcher et al.\ \cite{F04a}).
Other candidates for a dominating axisymmetric field are IC~342 (Krause et al.\
\cite{K89a}) and the LMC (Gaensler et al.\ \cite{G05}). A bisymmetric mode
dominates in M~81 (Krause et al.\ \cite{K89b}).

\emph{Magnetic arms\/} (Sect.~\ref{sect5}) may evolve between the optical
arms if the dynamo number is smaller in the gaseous arms than between
them, e.g. due to increased turbulent velocity of the gas in the arms
(Moss\ \cite{M98}, Shukurov\ \cite{S98}) or if magnetic turbulent diffusion
is larger in the arms.
The magnetic arms in NGC~6946 can be described as a superposition of an
axisymmetric $m=0$ and a quadrisymmetric $m=2$ mode (Rohde et al.\ \cite{R99}).
The RMs are preferably positive in the northern and negative in the southern
half of the galaxy (Fig.~\ref{n6946rm}). However, RM fluctuations on a scale
of about 1~kpc are superimposed onto the large-scale coherent field, and the
amplitude of the fluctuations is larger than that of the coherent field.

The massive spiral M~51 is of special interest: While previous low-resolution
data indicated a mixture of axisymmetric and bisymmetric fields in the
disk and a dominating axisymmetric field
in the halo (Berkhuijsen et al.\ \cite{B97}), recent observations with
higher resolution show that Faraday rotation fluctuates on small scales
with amplitudes much larger than those of the large-scale modes
(Fletcher et al.\ \cite{F06}). This means that the anisotropic random field
is stronger than the coherent regular field, possibly due to shearing
and compressing in a strong density wave flow.

Several other spiral galaxies revealed a highly ordered spiral pattern in
the polarization ($\mathbf{B}$--) vectors, but small Faraday rotation,
so that anisotropic random fields seem to dominate over the coherent
regular ones. In the barred galaxies NGC~1097 and NGC~1365 (Fig.~\ref{n1365})
the high degrees of polarization along the bar are mostly due to compressed
random fields (Beck et al.\ \cite{BF05}). Ram pressure from the intracluster
gas or interaction between galaxies have a similar compressional effect,
e.g. in the ``Antennae'' (Chy\.zy \& Beck\ \cite{C04}).

For most of about 20 nearby galaxies for which RM data are available,
no dominating magnetic modes could be reliably determined (Beck \cite{B00}).
Either these galaxies host a mixture of dynamo modes which cannot be resolved
due to the large telescope beam and/or to the low signal-to-noise ratio.
The Square Kilometre Array (SKA) will dramatically improve the
quality of polarization data (Beck \& Gaensler\ \cite{BG04}) and allow
detailed tests of galactic dynamo models (Beck\ \cite{B06}).

\section{Magnetic field strength and large-scale structure in the Milky Way}
\label{sect7}

The strength of the total field $\langle B_\mathrm{t}\rangle$
is $6~\mu$G near the Sun and about $10~\mu$G in the inner Galaxy
(Berkhuijsen, in Beck\ \cite{B01}), assuming equipartition between
the energy densities of magnetic fields and cosmic rays.
In our Galaxy the accuracy of the equipartition assumption can be
tested, because we have independent information
about the local cosmic-ray energy density from in-situ measurements
and about their radial distribution from $\gamma$-ray data.
Combination with the radio synchrotron data yields a local
strength of the total field $\langle B_\mathrm{t}\rangle$ of $6~\mu$G
(Strong et al.\ \cite{S00}), the same value as derived from
energy equipartition.
The radial scale length of the equipartition field of $\simeq12$~kpc
is also similar to that in Strong et al.\ (\cite{S00}).
In the nonthermal filaments near the Galactic center the field strength
may reach several 100~$\mu$G (Reich\ \cite{R94}, Yusef-Zadeh et al.\ \cite{Y96}),
but the pervasive diffuse field is much weaker (Novak\ \cite{N04}),
probably 20--40~$\mu$G (LaRosa et al.\ \cite{L05}, scaled for a
proton/electron ratio of 100).

Synchrotron polarization observations in the local Galaxy
imply a ratio of regular to total field strengths of
$\langle B_{\mathrm{reg}}/B_\mathrm{t}\rangle \simeq0.6$ (Berkhuijsen\ \cite{B71},
Brouw \& Spoelstra\ \cite{B76}, Heiles\ \cite{H96}).
The total radio emission along the local spiral arm requires that
$\langle B_\mathrm{reg}/B_\mathrm{t}\rangle $=0.6--0.7
(Phillipps et al.\ \cite{P81}). For $\langle
B_\mathrm{t}\rangle=6\pm2\,\mu$G these results give $4\pm1\,\mu$G
for the local regular field component (including anisotropic random fields).

\begin{figure}[htb]
\center
\centerline{\includegraphics[bb = 49 319 537 634,width=10cm,clip=]{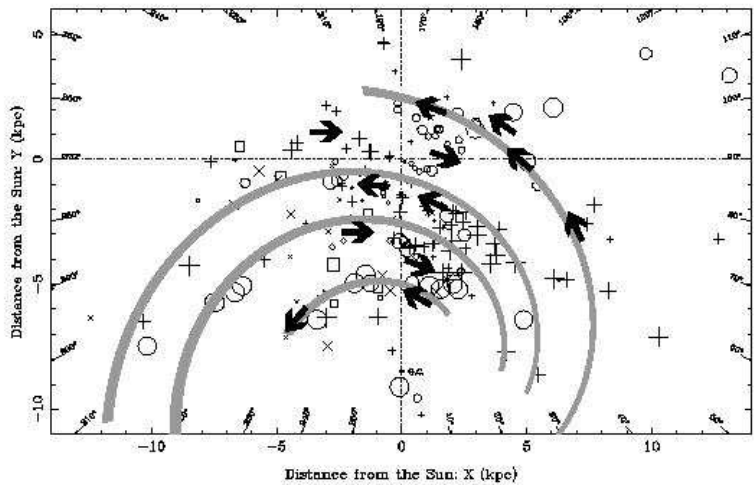}}
\vspace{0.2cm}
\caption{Bird'e eye view of the distribution of the RMs of pulsars
within $8^\circ$ of the Galactic plane. Positive RMs are shown as
crosses and X, negative RMs as circles and open squares.
The symbol sizes are proportional
to the square root of $|RM|$, with the limits of 5 and $\rm
250\,rad/m^2$. The directions of the bisymmetric field model are
given as arrows. The approximate location of four optical spiral
arms is indicated as dotted lines
(from \protect{Han et al.\ \cite{H99a}}).
}
\label{mfhan}
\end{figure}

Rotation measure and dispersion measure data of pulsars give an
average strength of the local coherent regular field of
$\langle B_{\mathrm{reg}}\rangle=1.4\pm0.2~\mu$G (Rand \& Lyne\ \cite{RL94},
Han \& Qiao\ \cite{H94}, Indrani \& Deshpande\ \cite{I98}).
In the inner Norma arm, the average strength of the coherent regular
field is $4.4\pm0.9~\mu$G (Han et al.\ \cite{H02}).
Han et al. (\cite{HM06}) claim that the regular field is stronger in the
Galactic arms than in interarm regions, in contrast to the results in
external galaxies (Sect.~\ref{sect3}). The values for
$\langle B_\mathrm{reg}\rangle$ derived from pulsar data are smaller than
the equipartition estimates. The former are underestimates if
small-scale fluctuations in field strength
and in electron density are anticorrelated, as expected for local
pressure equilibrium (Beck et al.\ \cite{B03}).
On the other hand, the equipartition values for $\langle B_\mathrm{reg}\rangle$
derived from polarized intensities overestimate the strength of the
coherent regular field if anisotropic turbulent fields (see Sect.~\ref{sect6})
are present.

The local Galactic field is oriented mainly parallel to the plane, with only a
weak vertical component of $B_z\simeq0.2~\mu$G (Han \& Qiao\ \cite{H94}).
Fig.~\ref{mfhan} shows the location of pulsars with measured RMs within the
Galactic plane and the derived directions of the large-scale field
(Han et al.\ \cite{H99a}). The Sun is located between two spiral arms, the
Sagittarius/Carina and the Perseus arms. The mean pitch angle of the spiral arms
is $\simeq -18^{\circ}$ for the stars and $-13^{\circ}\pm1^{\circ}$ for
all gas components (Vall{\'e}e\ \cite{V95}, \cite{V02}). Starlight polarization
and pulsar RM data (Fig.~\ref{mfhan}) give a significantly smaller pitch angle
($-8^{\circ}\pm1^{\circ}$) for the local magnetic field
(Heiles\ \cite{H96}, Han \& Qiao\ \cite{H94}, Indrani \& Deshpande\ \cite{I98},
Han et al.\ \cite{H99a}). The local field may form a \emph{magnetic arm\/}
located between two optical arms. Differences between the pitch angles of the
field and of the adjacent spiral arms of 10$^\circ$--20$^\circ$ were also
found in the spiral galaxy M~51 (Patrikeev et al.\ \cite{P06}).

\section{Magnetic field reversals in the Milky Way}
\label{sect8}

A puzzling property of the Galactic magnetic field is the existence of
several {\it reversals} along Galactic radius, derived from pulsar RM data
(Rand \& Lyne\ \cite{RL94}, Vall{\'e}e\ \cite{V95}, Han et al.\ \cite{H99a},
Frick et al.\ \cite{F01a}). To account for these reversals, an axisymmetric
field with radial reversals (Vall\'ee\ \cite{V96}) or a
\emph{bisymmetric\/} magnetic spiral with a small pitch angle have been
proposed (Han \& Qiao\ \cite{H94}, Indrani \& Deshpande\ \cite{I98},
Han et al.\ \cite{H02}, Fig.~\ref{mfhan}). Han et al. (\cite{HM06})
even claim that the interarm field is directed opposite to that in the arms.

It is striking that only very few field reversals have been detected in
external spiral galaxies. High-resolution maps of Faraday rotation, which measure
the RMs of the diffuse polarized synchrotron emission, are
available for a couple of spiral galaxies. A dominating bisymmetric field
structure was found only in M\,81 (Krause et al. 1989), but the pitch
angle is large and only two azimuthal reversals occur on opposite sides of the disk.
The disk fields of M\,51 and NGC~4414 can be described by a mixture of
dynamo modes (Berkhuijsen et al.\ \cite{B97}, Soida et al.\ \cite{S02}),
which appears as one single radial reversal to an observer
located within the disk (Shukurov\ \cite{S05}).
In NGC~2997, a radial field reversal between the disk and the central region
occurs at about 2~kpc radius (Han et al.\ \cite{H99b}).
However, no multiple reversals along radius, like those in the
Milky Way, were found so far in any external galaxy.

The discrepancy between Galactic and extragalactic data may be due to the
different observational methods. RMs in external galaxies are averages over the
line of sight through the whole disk and halo and over the large telescope beam,
and they may miss field reversals if these are restricted to a thin disk near
the galaxy plane. The results in the Milky Way are based on RMs of pulsars,
which trace the warm magneto-ionic medium near the plane along narrow lines of sight.
Some of the Galactic reversals may not be of Galactic extent, but due to
local field distortions or loops of the anisotropic turbulent field.
Pulsar RMs around a star formation complex indeed revealed a field
distortion which may mimic the reversal claimed to exist in the direction
of the Perseus arm (Mitra et al.\ \cite{M03}). More RM data from Galactic pulsars
are needed to obtain a clearer picture of the large-scale field structure.
The SKA will detect about 20000 pulsars and obtain RMs from most
of these (Cordes et al.\ \cite{CK04}).

\begin{figure}[htb]
\center
\centerline{\includegraphics[bb = 0 0 295 270,width=7cm,clip=]{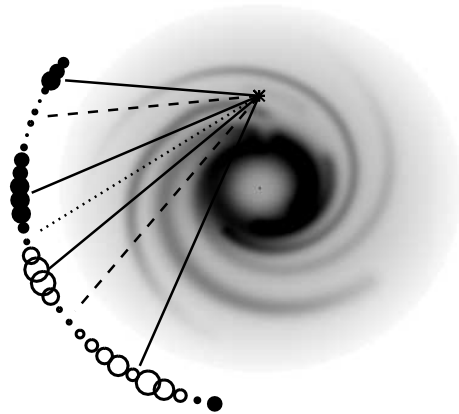}}
\vspace{0.2cm}
\caption{ Bird's eye view of the Galaxy, where the grey scale denotes
the electron density model by Cordes \& Lazio (\cite{CL03}) and circles
represent RMs of extragalactic sources, binned over 9$^\circ$ in Galactic longitude.
Open (closed) circles are negative (positive) RMs, and the largest circle denotes
RM$=-484$~rad~m$^{-2}$. The solid lines are $|$RM$|$ maxima, the dashed
lines are$|$RM$|$ minima, and the dotted line denotes a
large-scale sign change in RM (from \protect{Brown et al.\ \cite{B+06}}).
}
\label{rmbrown}
\end{figure}

Further indication that the structure of the local Galactic magnetic field is
rather complicated on pc and sub-pc scales
comes from RM maps obtained from the diffuse Galactic
synchrotron on which many reversals on small scales observed around 330~MHz
with the WSRT (Haverkorn et al.\ \cite{H03a}, \cite{H03b}). Major progress can
be expected from the new 1.4~GHz International Galactic Plane Survey observed
at the DRAO and ATCA telescopes (Brown \& Taylor\ \cite{BT01},
Haverkorn et al.\ \cite{H06}, Brown et al.\ \cite{B+06}).
It will provide high-resolution RM data of the
diffuse emission and several 100 RM values of polarized background sources.
Preliminary results of average extragalactic RMs (Fig.~\ref{rmbrown}) show
a smooth large-scale structure on scales of tens of degrees, obviously caused
by regular magnetic fields along the spiral arms. However, this appears to be
different from the results for external galaxies where the largest $|$RM$|$ are
found in interarm regions (Sect.~\ref{sect3}). RMs in Fig.~\ref{rmbrown} are
generally positive for Galactic longitudes smaller than 300$^\circ$ and negative
at larger longitudes, different from what one would expect from pulsar
RMs (Fig.~\ref{mfhan}). The decrease of $|$RM$|$ to almost zero (dashed and
dotted lines) can only occur if there are large-scale magnetic field reversals
along the line of sight, although not necessarily on Galactic scales.

Fig.~\ref{n6946rm} may help to explain some of the apparent discrepancies between
RM data from external galaxies and those in the Milky Way. An observer inside
NGC~6946 would detect the RM pattern of the weak coherent field only
for sufficiently long pathlengths (though along \emph{interarm\/} regions).
At distances below a few kpc, the
fluctuations in RM with frequent reversals dominate. We need a realistic
model of the field to predict what an internal observer would see, given a
limited number of pathlengths to pulsars or background sources.

\section{Small-scale field structures in the Milky Way}
\label{sect9}

Present-day polarization observations of external galaxies are restricted
to angular resolutions of $5^{\prime\prime}$--$10^{\prime\prime}$ because
at higher resolutions polarized intensities are too low. Before much more
sensitive telescopes like the SKA become operational, the small-scale structure
of interstellar magnetic fields has to be investigated by polarization observations
in the Milky Way.

The Galactic magnetic field has a significant turbulent component with a mean
strength of $\simeq5~\mu$G (Rand \& Kulkarni\ \cite{R89}, Ohno \& Shibata\ \cite{O93}).
Magnetic turbulence occurs over a large spectrum of scales, with the largest scale
determined from pulsar RMs of $l_\mathrm{turb} \simeq 55$~pc (Rand \& Kulkarni\
\cite{R89}) or $l_\mathrm{turb} \simeq 10 - 100$~pc (Ohno \& Shibata\ \cite{O93}).
These values are consistent with $l_\mathrm{turb} \simeq 20$~pc derived from
depolarization data in the galaxy NGC~6946 (Beck et al.\ \cite{B99b}).
In a test region in the southern Galactic plane, Haverkorn et al. (\cite{H04b})
determined the structure function of RM fluctuations which reveals an outer
scale of about 2~pc, possibly induced by HII regions.

\begin{figure}[htb]
\center
\centerline{\includegraphics[bb = 20 20 300 772,angle=270,width=12cm,clip=]{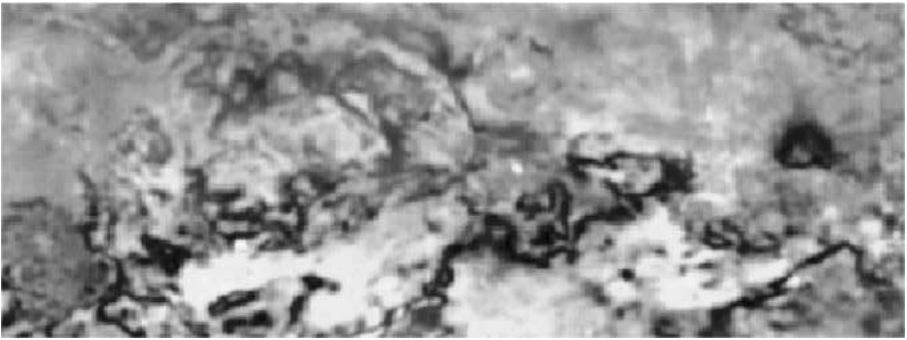}}
\vspace{0.2cm}
\caption{Polarized intensity around the plane of the Milky Way ($l=150^\circ -
174^\circ,\ b=-4.5^\circ - +4.5^\circ$) at 1.4~GHz ($\lambda21$~cm),
combined from Effelsberg and Dwingeloo data
(from \protect{Reich et al.\ \cite{R04a}}).
}
\label{canals}
\end{figure}

Major progress in detecting small structures has been achieved with decimeter-wave
observations of diffuse polarized emission from the Milky Way with the Effelsberg,
ATCA, DRAO, and WSRT telelscopes (Junkes et al.\ \cite{J87}, Duncan et al.\ \cite{D97},
\cite{D99}, Uyan{\i}ker et al.\ \cite{U98}, \cite{U99}, \cite{U03},
Gaensler et al.\ \cite{G01}, Uyan{\i}ker \& Landecker\ \cite{U02},
Haverkorn et al.\ \cite{H03a}, \cite{H03b}, \cite{H04a}, Kothes \& Landecker\
\cite{KL04}, Reich et al.\ \cite{R04a}). A wealth of structures
on pc and sub-pc scales has been discovered: filaments, canals, lenses, and
rings (Fig.~\ref{canals}). Their common property is to appear
only in the maps of polarized intensity, but not in total intensity. The
interpretation is hampered by several difficulties. Firstly, large-scale
emission in Stokes parameters Q and U is missing in interferometric and
even in single-dish maps so that the polarized intensities and angles
can be distorted severely (Reich et al.\ \cite{R04a}). Secondly, the wavelengths
of these polarization surveys are rather long, so that strong depolarization
of background emission in the foregound Faraday screen may lead to apparent
structures like {\it Faraday ghosts} (Shukurov \& Berkhuijsen\ \cite{S03}).
On the other hand, such features carry valuable information about the turbulent
ISM in the Faraday screen.

Another effect of Faraday depolarization is that, especially at decimeter
wavelengths, only emission from nearby regions may be detected.
The ISM is not always transparent for polarized radio waves, and the
opacity varies strongly with wavelength and position. The wavelength
dependence of Faraday depolarization allows \emph{Faraday tomography}
of different layers if maps at different (nearby) wavelengths are combined
(applying the method of \emph{RM synthesis} by Brentjens \& de Bruyn\
\cite{BB05}).

A Galactic 5~GHz polarization survey, which should be mostly free of Faraday
effects, has been started with the Urumqi telescope (Han \& Reich, in prep.).

\section{Magnetic fields in halos of galaxies and the Milky Way}
\label{sect10}

Most edge-on galaxies possess thick radio disks
of 1--3~kpc scale height (Lisenfeld et al.\ \cite{L04}) with
magnetic field orientations mainly parallel
to the disk (Dumke et al.\ \cite{D95}). A prominent exception is
NGC~4631 with the brightest and largest halo observed so far,
composed of vertical magnetic spurs connected to
star-forming regions in the disk (Golla \& Hummel\ \cite{G94}).
NGC~5775 is an intermediate case with parallel and vertical
field components (T\"ullmann et al.\ \cite{T00}, Fig.~\ref{n5775}),
similar to the Milky Way (see below).

\begin{figure}[htb]
\center
\centerline{\includegraphics[bb = 41 79 552 552,width=8cm,clip=]{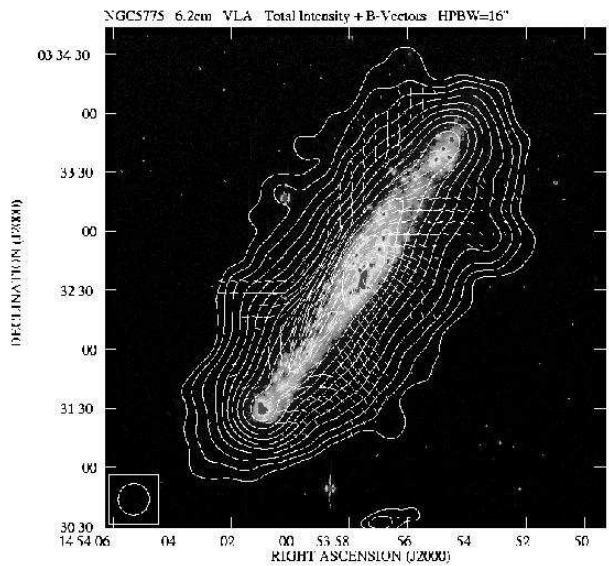}}
\vspace{0.2cm}
\caption{Total radio intensity (contours) and $\mathbf{B}$--vectors
of NGC~5775 at $16^{\prime\prime}$ resolution, observed with the
VLA at $\lambda$6~cm (from \protect{T\"ullmann et al.\ \cite{T00}}).
}
\label{n5775}
\end{figure}

The vertical full equivalent thickness of the thick radio disk of
the Milky Way is $\simeq3$~kpc near the
Sun (Beuermann et al.\ \cite{B85}, scaled to a distance to the
Galactic center of 8.5~kpc), corresponding to an exponential
scale height of $h_\mathrm{syn}\simeq1.5$~kpc. In case of equipartition
between the energy densities of magnetic field and cosmic rays,
the scale height of the total field is $\simeq 4$ times larger than
that of the synchrotron disk, $h_{B_\mathrm{t}} \simeq 6$~kpc.
$h_{B_\mathrm{t}}$ may be larger if cosmic rays originate
from star-forming regions in the plane and are not re-accelerated in the
halo, so that the electrons lose their energy above some height and the
equipartition formula yields too small values for the field strength
(Beck \& Krause\ \cite{BK05}).

Dynamo models predict the preferred generation of quadrupole fields
where the toroidal component (traced by RMs) has the same sign above
and below the plane. In external galaxies, no RM data of sufficient
quality are presently available. In the Milky Way, RMs of extragalactic
sources (see Han, in Wielebinski\ \cite{W05}, for a recent compilation)
and RMs of pulsars reveal that no large-scale reversal exists near
the plane for Galactic longitudes $\,l=90^{\circ} - 270^{\circ}$.
Thus the local field is part of a large-scale symmetric (quadrupole)
field structure parallel to the Galactic plane.
Towards the inner Galaxy ($l=270^{\circ} - 90^{\circ}$)
the signs are opposite above and below the plane. This may indicate
a global antisymmetric (dipole) mode (Han et al.\ \cite{H97},
Men \& Han\ \cite{MH03}) with
a poloidal field component perpendicular to the plane, but an
effect of local structures cannot be excluded. Remarkably,
submm polarization observations indicate a poloidal magnetic
field in low-density clouds in the vicinity of the Galactic center
(Chuss et al.\ \cite{CD03}).

\section{Concluding remarks}
\label{sect11}

Understanding and modelling the diffuse polarized emission from
the Milky Way needs major efforts from both the observational and
theoretical sides. At low frequencies, polarization data are of
limited use for CMB studies because they are seriously be affected by
Faraday effects. These, on the other hand, have the potential
to reveal new information on the structure and composition of the
magneto-ionic interstellar medium. For CMB studies and modelling,
mapping the Galaxy in polarization at high frequencies is required.
However, a polarization survey can only be reliably interpreted
if missing large-scale structures in Stokes Q and U, e.g. by missing
short spacings of synthesis telescopes, are corrected. External
galaxies are useful for modelling
the distribution of cosmic rays and the strength and large-scale
structure of interstellar magnetic fields. Results which are apparently
unique for the Milky Way, like frequent large-scale magnetic field
reversals and large $|$RM$|$ along spiral arms, should be taken with
caution and further investigated with better data.

\end{document}